\documentclass[twocolumn,superscriptaddress,showpacs]{revtex4}
\usepackage{graphicx}

\begin{document}
\title{Graphene nonlinearity unleashing at lasing threshold in graphene-assisted cavities}

\author{Alessandro Ciattoni}
\affiliation{Consiglio Nazionale delle Ricerche, CNR-SPIN, Via Vetoio, 67100 Coppito L'Aquila, Italy}

\author{Carlo Rizza}
\affiliation{Dipartimento di Scienza e Alta Tecnologia, Universit\`a dell'Insubria, Via Valleggio, 22100 Como, Italy} \affiliation{Consiglio Nazionale delle Ricerche, CNR-SPIN, Via Vetoio,  67100 Coppito L'Aquila, Italy}

\begin{abstract}
We investigate the nonlinear optical features of a graphene sheet embedded in an active cavity and we show that, when tuned near its lasing threshold, the cavity is able to isolate the spatially localized graphene nonlinearity thus producing a very strong nonlinear device response with multi-valued features. As opposed to standard situations where the small thickness of the graphene sheet hampers its remarkable nonlinear optical properties to be exploited, in our scheme the strong nonlinear optical regime is mainly triggered by the very intrinsic planar localization of graphene nonlinearity. The proposed strategy for exploiting graphene nonlinearity through its unleashing could open novel routes for conceiving ultra-efficient nonlinear photonic devices.
\end{abstract}


\maketitle

\section{Introduction}
Achieving versatile and externally driven light manipulation at chip-scale is a basic task of modern photonics since it would provide those optoelectronic circuit components (e.g. optical transistors, logic gates, etc.) required for devising a novel generation of ultra-high speed computing devices. After its efficient production through exfoliation, graphene has soon been identified as an excellent optoelectronic material \cite{Bonacc,BaoLoh} since the Dirac cone characterizing its electronic band structures provides a large carrier mobility yielding broadband and very efficient light-coupling \cite{Staube}. Besides chemical potential strongly affects graphene electron dynamics so that the optical response can be both structurally designed through chemical doping and rapidly driven through externally applied bias voltages. Accordingly a number of graphene-based devices have been proposed as polarizers \cite{BaoZh1,ChengC,Bludov}, optical modulators \cite{LiuYin,Sensal,Goscin,GanShi,Majumd}, photodetectors \cite{ParkAh,XiaMue,Muelle,Wither}, saturable absorbers and mode-locked ultrafast lasers \cite{BaoZh2,XingGu,SunHas}. In addition, the linear band structure provides graphene with a large and broadband Kerr-like optical nonlinearity since at each photon frequency there is an available interband optical transition \cite{Mikha1,Mikha2,Hendry,Ishika,ZhangV,ChenWa}. Accordingly, a number of remarkable effects due to such pronounced nonlinearity have been considered, such as efficient third harmonic generation \cite{HongDa}, nonlinear surface plasmons propagation \cite{Gorba1}, propagation of nonlinear modes through dielectric waveguides hosting graphene sheets \cite{Audito} and nonlinear optical propagation in graphene-clad tapered fibers \cite{Gorba2}. Other relevant nonlinear phenomena which have been predicted to be supported by graphene are optical bistability in the terahertz range \cite{Peress}, propagation of subwavelength optical solitons \cite{Nester} and propagation of discrete solitons in graphene metamaterials \cite{Bludov}.

Even though remarkable, such achievements have the common drawback of requiring relatively large optical powers as a consequence of the absolute smallness of the graphene nonlinear susceptibility (which is yet much larger than that of insulating materials \cite{Hendry}). Therefore, as in standard nonlinear optics without graphene, a field enhancement mechanism has to support the graphene-based nonlinear setup to trigger a low-intensity strong nonlinear optical regime. Recently, strong field enhancement effects have been considered in the presence of graphene \cite{GanMak,Thongr}. However, to the best of our knowledge, achieving feasible low-intensity nonlinear light steering by combining graphene optical nonlinearity with a field enhancement mechanism has been shown in a single paper by Gu {\it et al.} \cite{GuPetr} where the authors show that placing a graphene sheet on the top of a silicon photonic crystal hosting a high-Q cavity (responsible for a large in-cavity field enhancement) produces optical bistability, self induced regenerative oscillations and coherent four-wave mixing at ultra-low optical intensities.

In this paper, we show that a graphene sheet, placed within an active cavity tuned near its lasing threshold, turns the linear cavity behavior into a highly nonlinear one, transmissivity and reflectivity being multi-valued functions of both optical intensity and graphene chemical potential. Recently, it has been proposed to use graphene sheets within passive optical cavities for improving matter-radiation control \cite{Furchi,EngelS} and within active optical cavities for achieving highly efficient second harmonic generation \cite{Ciatto}. On the other hand it is well known that within an active cavity at its lasing threshold the gain supported by the inverted active medium fully compensates cavity losses and radiation leakage so that the cavity field undergoes a full replication after a round trip \cite{Svelto}. Here we consider a novel mechanism where the cavity at its lasing threshold literally unleashes the graphene nonlinearity since, in this situation, the field replication after a round trip inside the cavity amounts to a linear propagation compensation which, in turn, leaves graphene nonlinearity as the main agent ruling the cavity behavior.

Generally, harnessing graphene nonlinearity in a bulk device is hampered by its intrinsic planar localization whereas the strategy discussed in this paper overcomes the difficulty just exploiting localization. Since the proposed mechanism operates at low optical intensities and is supported by a simple setup, we believe it can pave the way for a novel generation of compact nonlinear light-steering devices.

The paper is organized as follows. In Section II we analyze the nonlinear optical response of an active cavity hosting a graphene sheet. In Section III we analyze the mechanism allowing the cavity lasing threshold to unleash graphene nonlinearity and we conclude that their combination triggers a strong nonlinear optical regime. In Section IV we discuss the feasibility of the considered nonlinear regime through a numerical example. In Section V we draw our conclusions.

\begin{figure}
\includegraphics[width=0.4\textwidth]{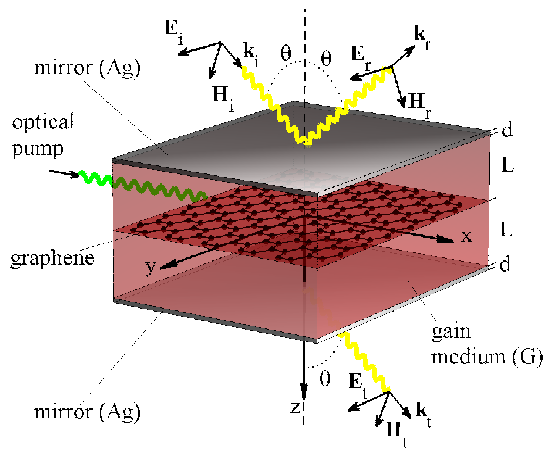}
\caption{(Color online) Geometry of the graphene layer embedded within a gain medium (G) enclosed by two metallic (Ag) mirrors. The cavity is illuminated by an inclined monochromatic wave (i) which produces reflected (r) and transmitted waves (t). The cavity is also illuminated by an inverting pump beam.}
\end{figure}

\section{Optical response of a graphene-assisted cavity}

In Fig.1, the graphene assisted cavity is sketched together with the geometry of the interacting electromagnetic field. The graphene sheet is embedded between two layers (G) of thickness $L$ filled by a gain medium (e.g. a dye-doped polymer) enclosed
by two metallic mirrors (Ag) of thickness $d$. The cavity is excited by a monochromatic inclined TE-polarized plane wave (i) producing reflected (r) and transmitted (t) fields. We have chosen TE polarization to avoid complications arising from the longitudinal electric field component within the cavity. In our analysis, we neglect the gain nonlinearity by self-consistently assuming the optical intensity within the bulk to be much smaller than the gain saturation intensity. As a consequence, propagation of laser radiation within the active medium (and through the mirrors) is purely linear, the free cavity (without graphene) being a standard linear optical device. Due to the extremely small thickness of the graphene layer, we model
its effect through the matching conditions $E_y^+ - E_y^- = 0$ and $H_x^+ - H_x^- = K_y$ i.e. the continuity of the electric field tangential component and the discontinuity of the magnetic field tangential component produced by the graphene surface current \cite{Mikha2}
\begin{equation} \label{surface}
K_y = \sigma_1 E_y + \sigma_3 |E_y|^2 E_y
\end{equation}
where $E_y=E_y^+=E_y^-$. Here the nonlinear corrections to the surface current have been taken into account up to the first order and the effect of higher harmonics generation have been neglected (see Appendix A). Both the linear surface conductivity $\sigma_1$ and its nonlinear correction $\sigma_3$ are strongly affected by the graphene chemical potential thus providing the system an overall tunability through chemical doping and electrical gating. Exploiting the spatial localization of the graphene nonlinearity we obtain the equation (see Appendix B)
\begin{equation} \label{BasicEq}
\left[ \frac{1}{2} |\Lambda|^2 \sqrt{\frac{\mu_0}{\epsilon_0}} \sigma_3 |E_t|^2 +\left(\frac{\Omega}{\Lambda} + \frac{1}{2} c \mu_0 \sigma_1 \right)
\right] E_t = \frac{\cos \theta}{\Lambda^2} E_i
\end{equation}
relating the amplitudes $E_i$ and $E_t$ of the incident and transmitted waves, respectively. Here $\Omega$ and $\Lambda$ are dimensionless complex factors depending on the radiation frequency, the inclination angle $\theta$, and the gain medium and mirrors' slabs thicknesses $L$,$d$, and dielectric permittivities, $\epsilon_G$ and $\epsilon_{Ag}$. Note that, if the nonlinear contribution to the surface current of Eq.(\ref{surface}) is negligible ($\sigma_3=0$), Eq.(\ref{BasicEq}) yields
\begin{equation} \label{linear}
\frac{|E_t|^2}{|E_i|^2} = \frac{\cos^2 \theta}{|\Lambda|^4 \left|\frac{\Omega}{\Lambda} + \frac{1}{2} c \mu_0 \sigma_1 \right|^2}
\end{equation}
which is the transmittance function of the linear cavity.

\section{Mechanism supporting the strong nonlinear regime}

In order to investigate the transmittance $T = |E_t|^2 / |E_i|^2$ of the graphene-assisted cavity, we introduce the dimensionless quantities
\begin{eqnarray} \label{parameter}
\widetilde{T} = \left( \frac{ |\Lambda|^4}{2\cos\theta} c \mu_0 |\sigma_3|  |E_i|^2 \right)^{2/3} T, \nonumber \\
W = \frac{|\sigma_3|}{\sigma_3}  \frac{2 |\Lambda|^{2/3} \left(\frac{\Omega}{\Lambda} + \frac{1}{2}  c \mu_0 \sigma_1 \right)} {\left( 3 \cos^2 \theta c
\mu_0 |\sigma_3| |E_i|^2 \right)^{1/3}}
\end{eqnarray}
since Eq.(\ref{BasicEq}), after taking the square modulus of both its members, yields
\begin{equation} \label{MainEq}
|\widetilde{T}+W|^2 \widetilde{T} = 1.
\end{equation}
This general equation equation has the details of the specific device and of the excitation absorbed in the parametrization of Eqs.(\ref{parameter}). Here $\widetilde{T}$ is a real quantity proportional to the transmittance $T$ and $W$ is a complex parameter depending on the system and excitation. In Fig.2a we plot the surface $\widetilde{T}$ on the complex plane $W$ as evaluated from Eq.(\ref{MainEq}) and we stress that it describes the overall optical response of any graphene-assisted cavity. Note that, due to the cubic term (in turn produced by the graphene nonlinearity), $\widetilde{T}$ is generally a multi-valued function of $W$, having three different values within the shadowed region of Fig.2a. In Fig.2b we plot a portion of the region $M$ (which is actually unbounded toward $Re(W) \rightarrow -\infty$) of the complex plane $W$ on which the system transmittance is multi-valued. Therefore, we conclude that the system supports a strong nonlinear regime with a multi-valued transmittance whenever it is excited in such a way that $W$ is in M.

\begin{figure}
\includegraphics[width=0.4\textwidth]{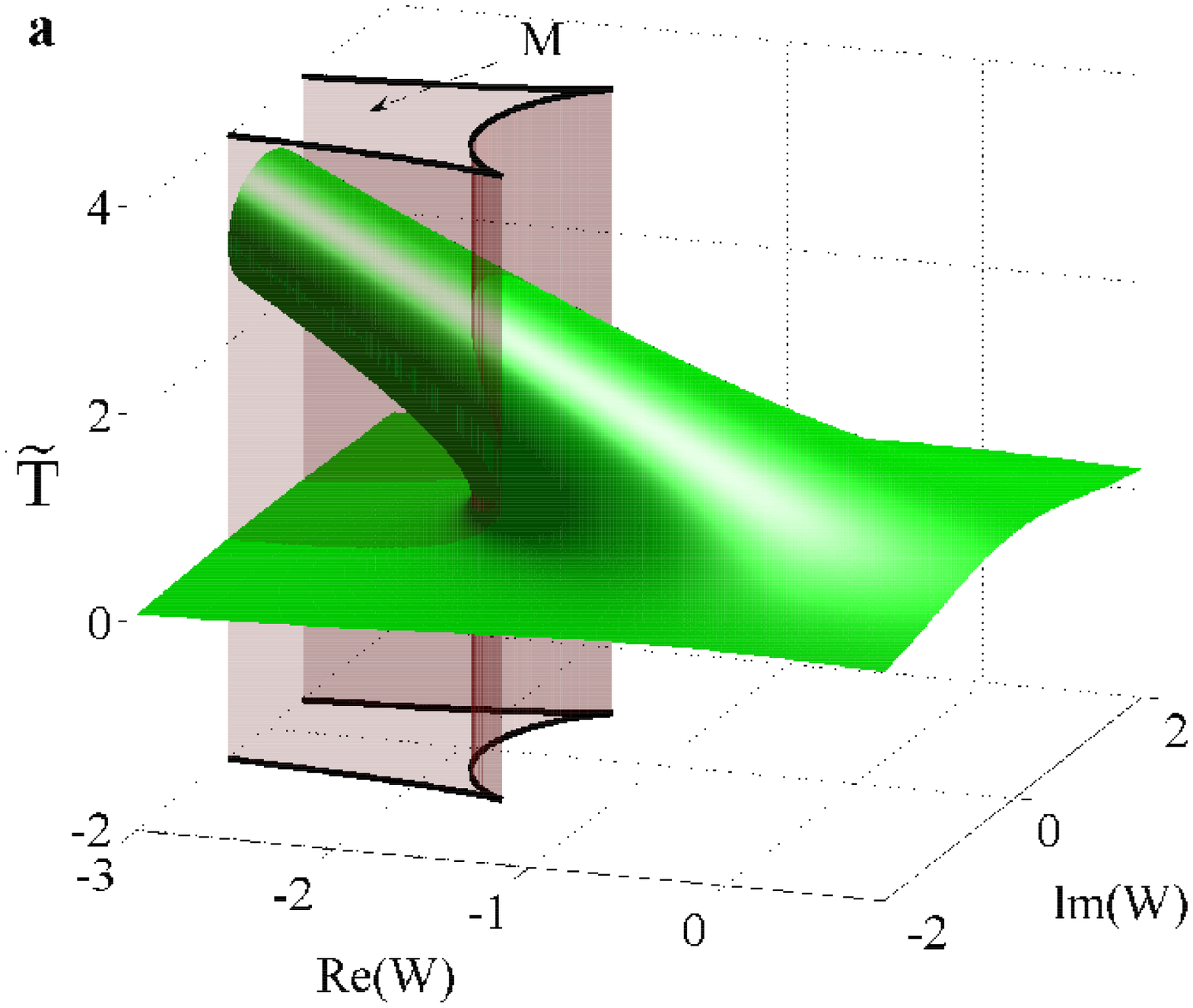}
\includegraphics[width=0.4\textwidth]{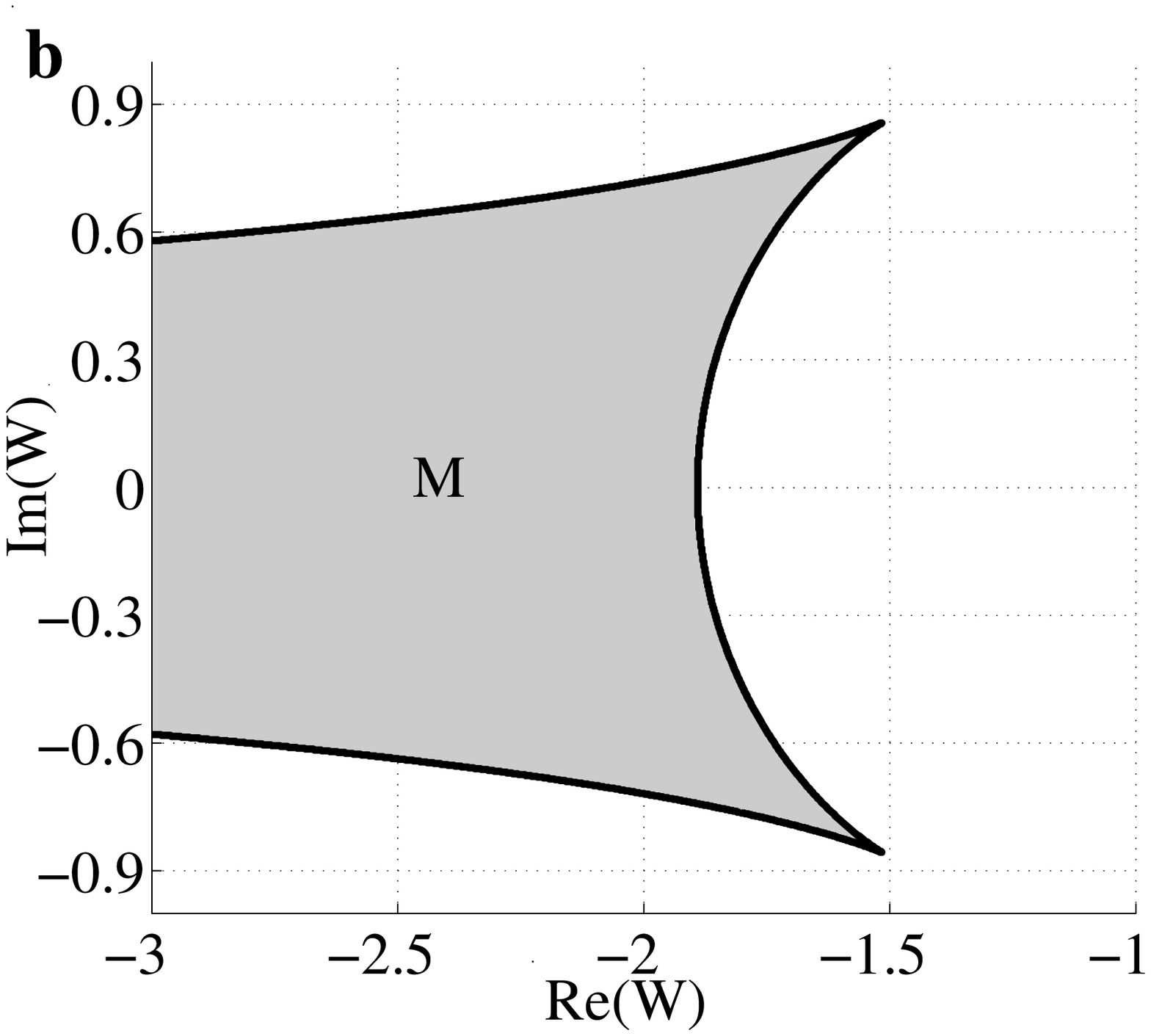}
\caption{(Color online) (a) Modified cavity transmissivity
$\widetilde{T}$ versus the complex parameter $W$ characterizing the cavity status and excitation, showing the occurrence of a strong nonlinear system behavior. Due to normalization, such surface characterizes any possible graphene-assisted cavity response. (b) Region $M$ of the complex plane $W$ where the transmissivity is multi-valued. As explained in the text this region physically corresponds to a cavity excited close to its lasing threshold.}
\end{figure}

The values of $W$ belonging to the region $M$ select the cavities actually hosting the predicted strong nonlinear regime and such region $M$ allows the underlying physical supporting mechanism to be highlighted. The key observation is that the absolute value $|W|$ on the region $M$ of Fig.2b is of the order of unity (at least in the relevant region around its right boundary) and this requirement has to be compared with the second of Eqs.(\ref{parameter}). Note that, as a consequence of the weakness of the nonlinear cubic term in Eq.(\ref{surface}), $c\mu_0 |\sigma_3| |E_i|^2 \ll 1$ unless the incident optical intensities is so large to make Eq.(\ref{surface}) useless and, at the same time, to severely damage the structure through heating. Therefore, in order to trigger the strong nonlinear regime at feasible optical intensities, from the second of Eqs.(\ref{parameter}) we conclude
that, the necessary requirement is
\begin{equation} \label{condition}
\Psi \equiv \left| \frac{\Omega}{\Lambda} + \frac{1}{2}  c \mu_0 \sigma_1 \right| \ll 1.
\end{equation}

Considering Eq.(\ref{linear}), we note that the condition in Eq.(\ref{condition}) would lead the transmissivity of the linear cavity to be greater than one and this can only occur if an energy supplying mechanism is present in the bulk, thus justifying our choice of embedding the graphene layer in a gain medium (see Fig.1a). Condition in Eq.(\ref{condition}) corresponds to an active cavity operating close to one of its lasing threshold and this clarifies the physical mechanism supporting the strong nonlinear regime. Indeed, close to the lasing threshold, the wave inside the cavity starting from the graphene plane ($z=0^+$) makes a complete round trip and returns to the graphene plane ($z=0^-$) with almost the same amplitude and phase. Such an
oscillation is equivalent to a linear propagation compensation of the field inside the cavity. Therefore the nonlinearity localized on the graphene plane, being the only residual agent not compensated by the cavity, can fully rule the field dynamics. We refer to this mechanism as the unleashing of the graphene nonlinearity produced by the cavity close to a lasing threshold. A different but equivalent way of grasping the same mechanism is observing that in Eq.(\ref{BasicEq}) the nonlinear term (containing $\sigma_3$) is generally negligible thus leading the cavity to exhibit the linear response of Eq.(\ref{linear}). However, if condition in Eq.(\ref{condition}) holds, the linear contribution in the LHS of Eq.(\ref{linear}) is very small as well so that the nonlinear term cannot be neglected with the result of producing a marked nonlinear behavior. It is worth stressing that such strong nonlineary cannot be observed without graphene since the free cavity is a purely linear device and the localization of the nonlinearity on a single plane plays, as explained, a fundamental role.

\begin{figure}
\includegraphics[width=0.4\textwidth]{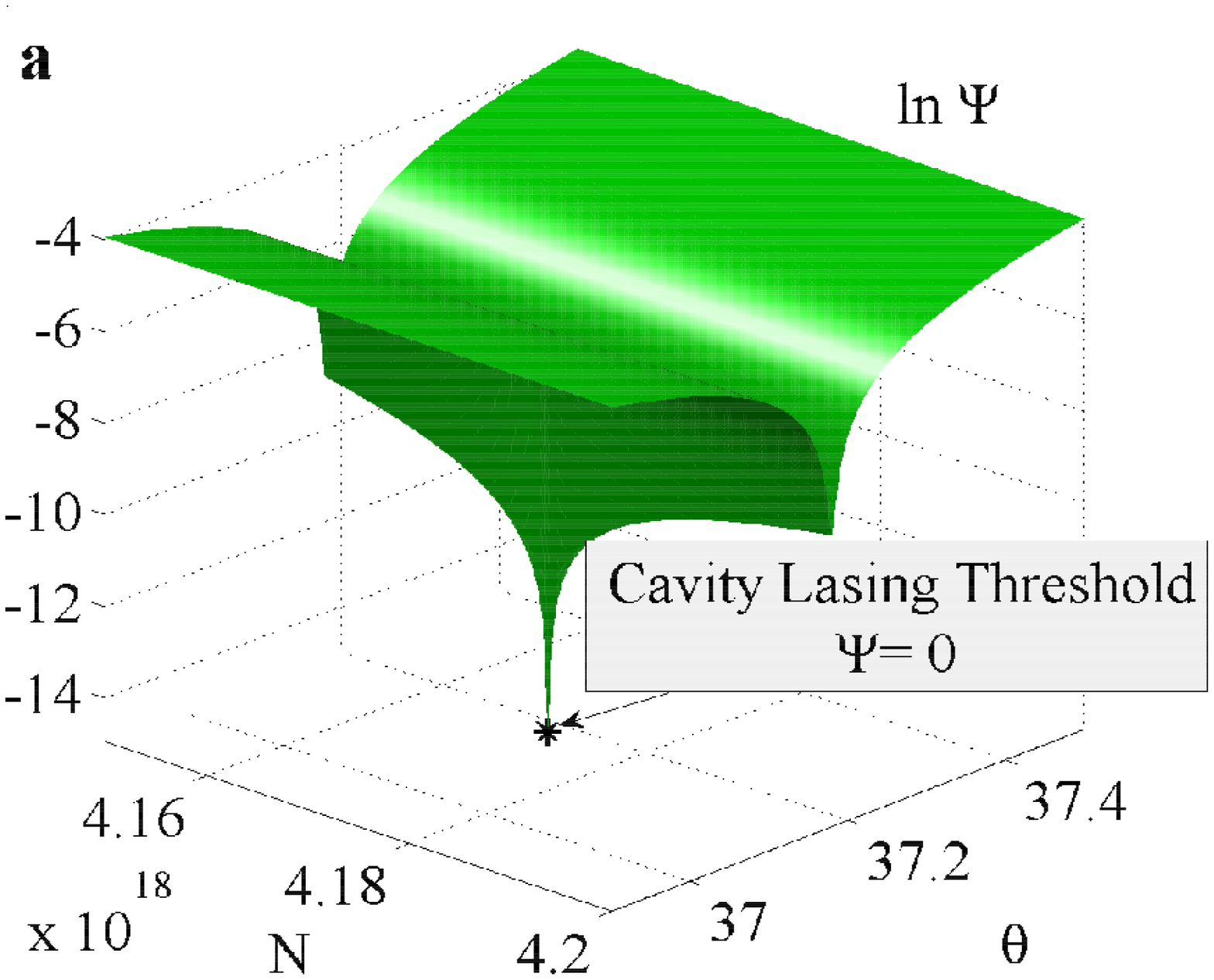}
\includegraphics[width=0.4\textwidth]{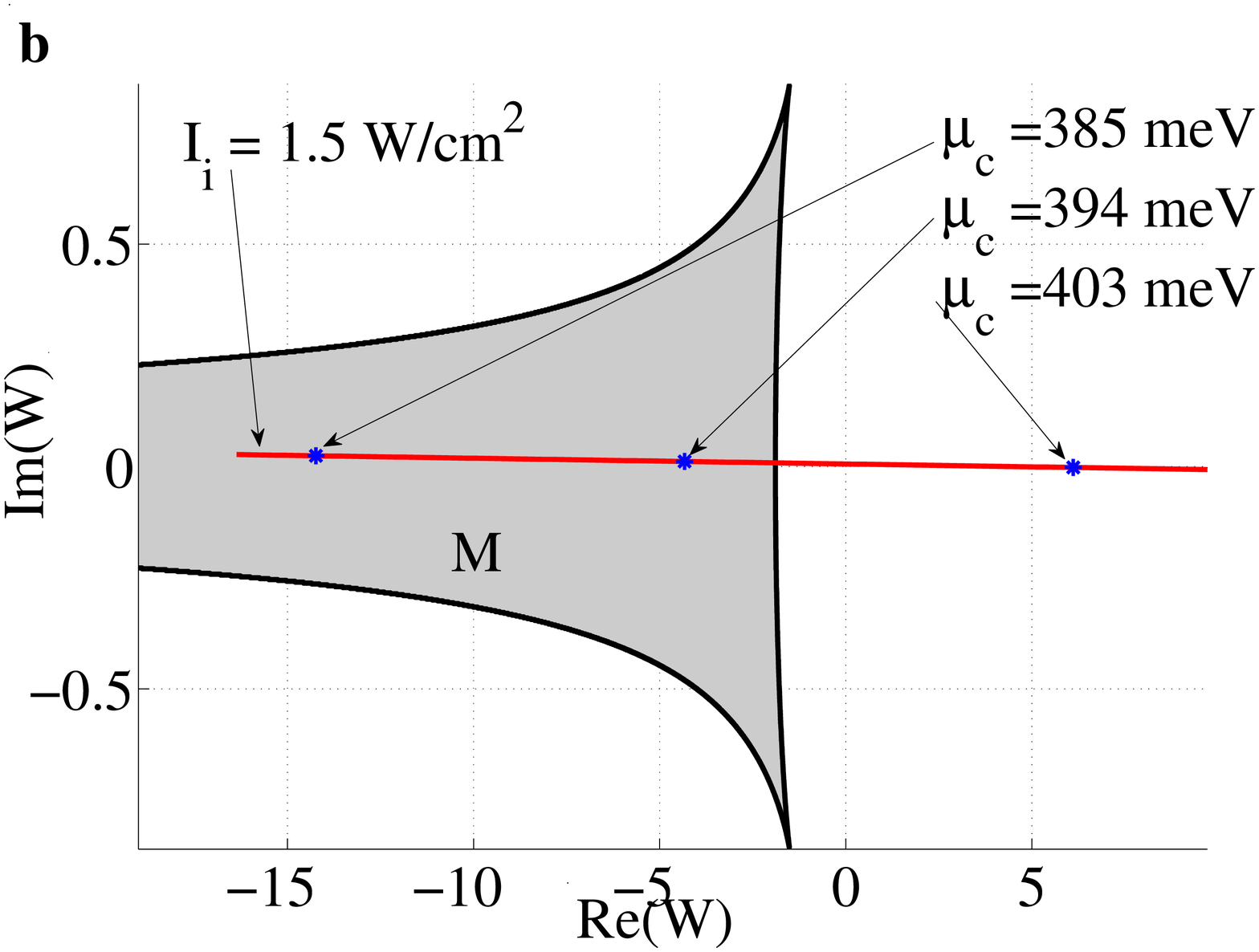}
\caption{(Color online) (a) Logarithmic plot of the function $\Psi$ whose zeros correspond to the cavity lasing thresholds. In this example all the cavity parameters are held fixed except for the incidence angle $\theta$ and the concentration of the dye molecules $N_T$. The sharp peak corresponds to the cavity threshold. (b) The oblique solid line represents the cavity states with fixed incident optical intensity $I_i = 1.5 \: W/cm^2$ and varying chemical potential $\mu_c$. The shadowed region $M$ is the same as in Fig.2b and it contains the states where the cavity response is multi-valued. The intersection of the solid line with $M$ proves that, for the considered optical intensity, the cavity actually shows a highly nonlinear behavior with multi-valued response features.}
\end{figure}

\section{Feasibility of graphene nonlinearity unleashing}

We now discuss the predicted strong nonlinear regime in a realistic situation at optical frequencies. We consider a cavity whose mirrors are silver layers of thickness $d=0.22 \: \mu m$ and whose gain slabs are filled by Rhodamine 6G (Rh6G)-doped polymethyl methacrylate (PMMA) of thicknesses $L=0.7 \: \mu m$. The cavity is pumped by an inverting laser beam at a wavelength of $532 \:
nm$. We have chosen $\lambda = 565 \: nm$ as the field wavelength since it is near to the gain peak \cite{SoanCa}. The silver and PMMA permittivities are, at the considered field wavelength, $\epsilon_{Ag} = -11.9641+0.8310i$ and $\epsilon_{PM} = 2.2282$, respectively. The permittivity of the gain medium is given by $\epsilon_G = \epsilon_{PM} - i \frac{\lambda}{2 \pi} \sqrt{\epsilon_{PM}} \frac{\sigma_e N_T}{1+I/I_{se}'} \frac{I_p/I_{sp}}{1+I_p/I_{sp}}$ \cite{SoanCa} where $I$ is the optical intensity within the medium bulk, $I_p$ is the pump intensity, $I_{sp}$ is the pump saturation intensity, $I_{se}'=I_{se}(1+I_p/I_{sp})$ (where $I_{se}=300 \: kW/cm^2$) is the field saturation intensity,  $\sigma_e = 1.2 \times 10^{-16} \: cm^2$ is the emission cross section and $N_T$ is concentration of the Rh6G molecules. We have chosen the strong pump saturation regime by setting $I_p/I_{sp} = 5000$ so that $\frac{I_p/I_{sp}}{1+I_p/I_{sp}} \simeq 1$ and $I_{se}' \simeq 1500 \: MW/cm^2$ which is a sufficiently high saturation intensity to self-consistently assume that $I/I_{se}' \ll 1$ (a condition we have {\it a-posteriori} checked, see below) and to express the permittivity of the gain medium as $\epsilon_G = \epsilon_{PM} - i \frac{\lambda}{2 \pi} \sqrt{\epsilon_{PM}} \sigma_e N_T$.

As discussed in the above paragraph, Eq.(\ref{condition}) is the necessary requirement for observing the strong nonlinear regime. We have set $\mu_c = 400 \: me V$ so that, using the expression for $\sigma_1$ (see Appendix A) and the above parameters, $\Psi$ turns out to be a function of the incidence angle $\theta$ and the Rh6G molecules concentration $N_T$ and the cavity lasing thresholds are the zeros of this function. In Fig.3a we draw the logarithmic plot of $\Psi$ around one of its zeros which is located at $\theta =  37.1783 \: deg$, $N_T = 4.1720 \times 10^{18} \: cm^{-3}$. Note that, even though the region where $\Psi \ll 1$ is rather sharp in Fig.3a, the cavity threshold can be simply experimentally achieved by using a beam with small angular
divergence around the predicted critical angle.

\begin{figure}
\includegraphics[width=0.4\textwidth]{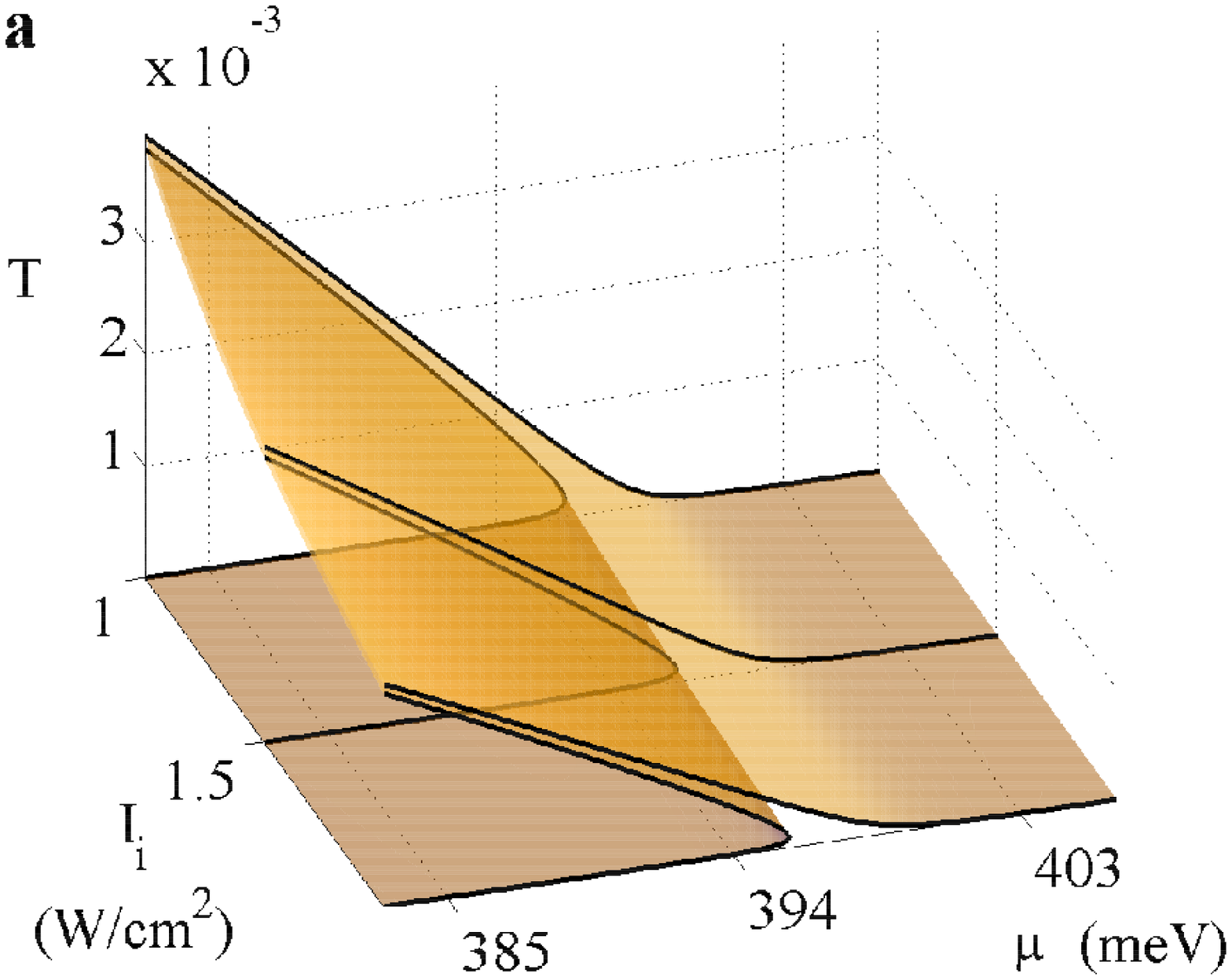}
\includegraphics[width=0.4\textwidth]{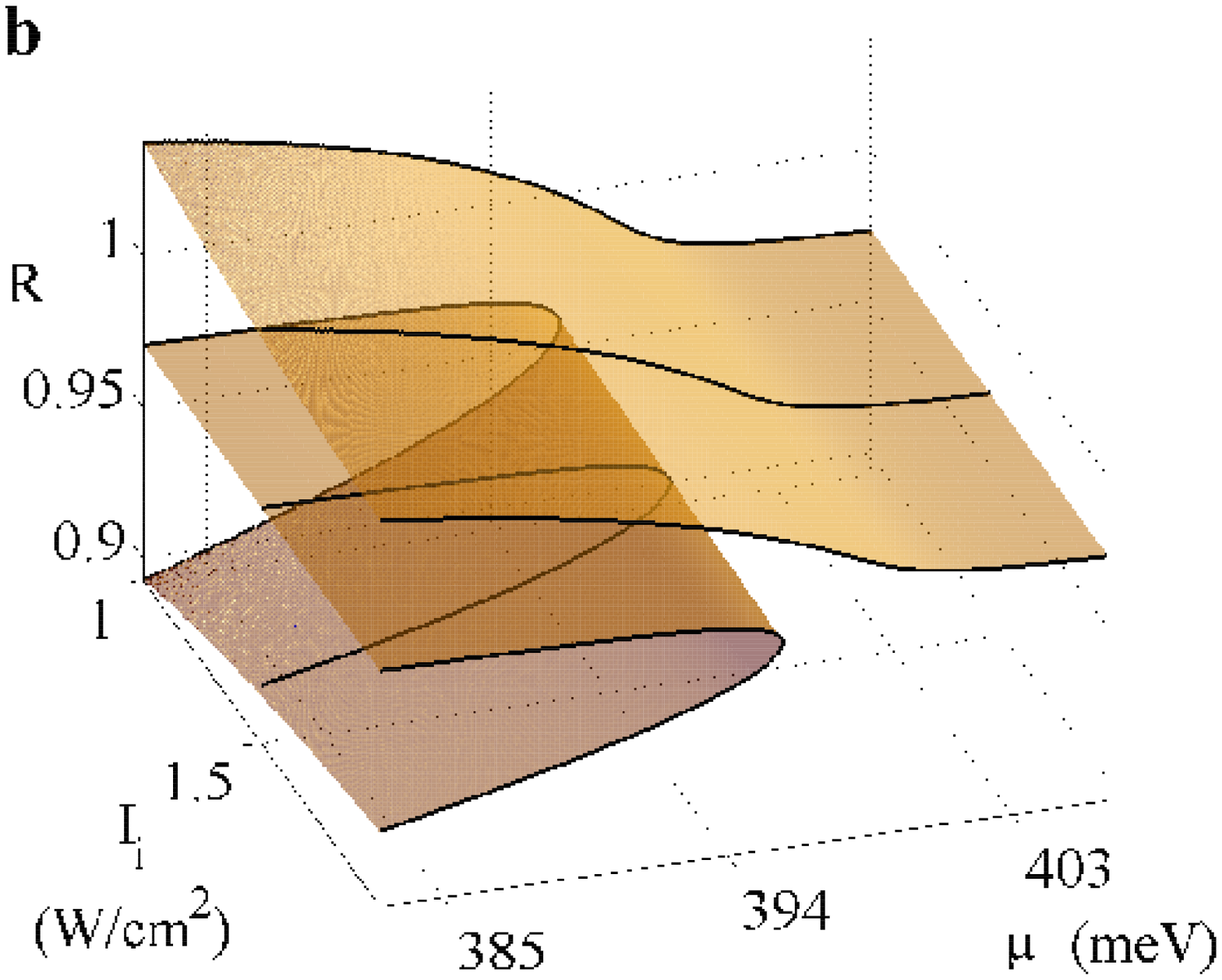}
\caption{(Color online) Cavity transmissivity $T$ (a) and reflectivity $R$ (b) plotted against the incident intensity $I_i$ and the chemical potential $\mu_c$.}
\end{figure}

By slightly detuning the cavity from the lasing threshold, the highly nonlinear multi-valued behavior of the transmissivity appears. With all the above parameters fixed (except the chemical potential), we have chosen to investigate the system response for various chemical potentials in the range $383 \: meV < \mu_c < 403 \: meV$ (which are values sufficiently close to the above used $\mu_c = 400 \: meV$) and various incident optical intensities $I_i = c \epsilon_0 |E_i|^2 /2$ spanning the range $1 \: W/cm^2 < I_i < 2 \: W/cm^2$. Using the numerical value $\sigma_3 = -i 8.16 \cdot 10^{-23} A m^2 V^{-3}$ (see Appendix A) for the graphene nonlinear conductivity, we have evaluated the complex $W$ parameters (see the second of Eqs.(\ref{parameter})) obtained by setting $I_i = 1.5 \: W/cm^2$ and varying $\mu_c$ and we have plotted them on the complex plane in Fig.3b as an oblique solid line. In the same Fig.3b we have also reported the shadowed region $M$ of Fig.2b. The intersection of the solid oblique line with the shadowed $M$ region proves that for the consider values of $\mu_c$ and $I_i$ the cavity actually shows a multi-valued behavior of the transmissivity.

In Figs.4a and 4b we have plotted the cavity transmissivity $T$ and reflectivity $R=|E_r|^2/|E_i|^2$ evaluated from Eq.(\ref{MainEq}) in the considered ranges of $\mu_c$ and $I_i$. Note the multi-valued structures of both $T$ and $R$ which are the key feature of the above discussed strong nonlinear regime.

\begin{figure}
\includegraphics[width=0.45\textwidth]{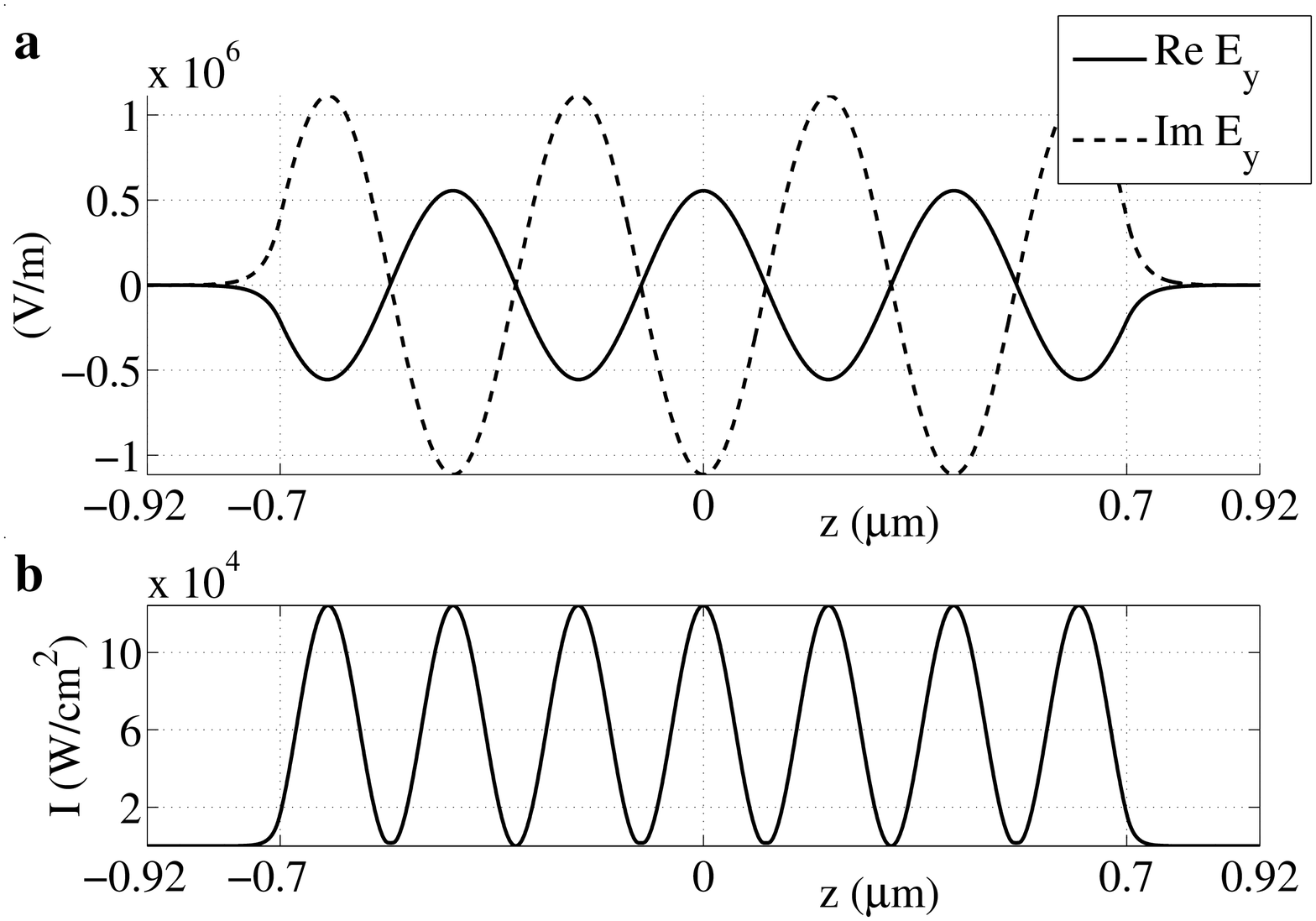}
\caption{(Color online) (a) Real and imaginary parts of the electric field $E_y$ within the cavity corresponding to the higher value of the transmissivity pertaining the excitation state $I_i = 1.5 W/cm^2$ and $\mu_c = 385 \: meV$. The field enhancement produced by the cavity is particularly evident since for the considered intensity, the incident electric field amplitude is $|E_i| = 3.36 \cdot 10^3 V/m$. Note that, since the electric field is practically left invariant by the reflection $z \rightarrow -z$, the field almost replicates itself after a complete round trip inside the cavity, the fundamental physical ingredient allowing the considered strong nonlinear regime to occur. (b) Optical intensity $I$ within the cavity corresponding to the situation of panel (a). Note that such optical intensity is much smaller than the considered saturation intensity $I_{se}' \simeq 1500 \: MW/cm^2$ so that the model for the gain medium we have used is self-consistently
correct.}
\end{figure}

In Fig.5a we plot the real and imaginary parts of the electric field $E_y$ within the cavity (see Appendix A) corresponding to the higher value of the transmissivity pertaining the excitation state $I_i = 1.5 \: W/cm^2$ and $\mu_c = 385 \: meV$. Note that, in addition to the evident field enhancement produced by the cavity (since for $I_i = 1.5 \: W/cm^2$ the incident electric field amplitude is $|E_i| = 3.36 \cdot 10^3 V/m$), the electric field is practically left invariant by the reflection $z \rightarrow -z$ and this confirms our above reasoning that the field almost replicates itself after a complete round trip inside the cavity thus allowing the graphene nonlinearity on the plane $z=0$ to strongly rule the cavity behavior. In Fig.5b we plot the optical intensity $I$ within the cavity (see Appendix B) which, being much smaller then the above considered saturation intensity $I_{se}' \simeq 1500 \: MW/cm^2$ self-consistently assures that the model for the gain medium we have used is correct.

\section{Conclusions}
A novel strategy has been proposed to fully exploit the graphene nonlinearity for achieving a very strong nonlinear regime. Specifically it has been suggested and theoretically proved that a standard active cavity hosting a graphene sheet, when excited close to its lasing threshold, is able to isolate the planar localized graphene nonlinearity as a consequence of the field replication after a round trip inside the cavity. Such a graphene nonlinearity unleashing turns the simple linear cavity into a highly nonlinear photonic device at very low optical intensity with a marked multi-valued trait. A specific example has been considered at optical frequencies where all the above predicted features have been checked. It should be stressed that such example also highlights the efficiency of the proposed mechanism since, as it is well known, graphene nonlinearity at optical frequencies is much weaker than the remarkable one appearing in far infrared and terahertz ranges (note that very few graphene-based nonlinear optical setups have been proposed in literature). On the other hand, since suitable gain mechanisms exist also at frequency lower than the optical ones, the proposed graphene nonlinearity unleashing can in principle be exploited at such lower frequencies where an even stronger nonlinear regime is therefore expected. Among the possible applications of the considered setup it is worth stressing that the predicted multivalued response of both cavity transmissivity and reflectivity can be exploited to conceive devices for optical information processing such as optoelectronic memory units operating at low optical intensity.

\section{Acknowledgements}
The authors thank the U.S. Army International Technology Center Atlantic for financial support (Grant No. W911NF-14-1-0315).
The authors gratefully acknowledge fruitful discussions with Dr. Eugenio Del Re.

\appendix
\section{Linear and nonlinear graphene conductivities}
The surface conductivity of the graphene sheet, if $k_B T \ll |\mu_c|$, can be expressed as \cite{Hanson}
\begin{eqnarray}
\sigma_1 &=& \frac{ ie^2 k_B T }{\pi \hbar^2 (\omega + i2\Gamma)}\left( \frac{\mu_c}{k_B T} + 2 \ln \left( e^{-\frac{\mu_c}{k_B T}} +1 \right) \right)
             \nonumber \\
         &+& \frac{i e^2}{4 \pi \hbar} \ln \left( \frac{2|\mu_c| - (\omega + i2\Gamma) \hbar} {2|\mu_c| + (\omega +i2\Gamma) \hbar } \right)
\end{eqnarray}
where $e$ is the electron charge, $\hbar$ and $k_B$ are Planck's and Boltzmann's constant, respectively, $T$ is the temperature which is here fixed at $300$ K, $\mu_c$ the graphene chemical potential and $\Gamma$ is a phenomenological scattering rate accounting for the graphene intrinsic losses which is here fixed at $\Gamma = 0.43$ meV. 

In order to have a realistic estimation of the nonlinear conductivity coefficient $\sigma_3$ we have resorted to measured numerical values reported in literatured. In Ref.\cite{ZhangV} the authors exploit the Z-scan technique to measure the nonlinear refractive index of graphene for the wavelength $\lambda = 1550 \: nm$ and they obtain $n_2 = 1.5 \cdot 10^{-9} \: cm^2 W^{-1}$ with negligible imaginary part. The corresponding nonlinear susceptibility is $\chi_3 = 8 \cdot 10^{-14} m^2 V^{-2}$ which, exploting the relation $\sigma_3 = -i \omega \epsilon_0 d \chi_3$ (where $d= 3.3 \cdot 10^{-10} \: m$ is the graphene layer thickness), yields $\sigma_3 (\lambda= 1550 \: nm) = -i 4.62 \cdot 10^{-21} A m^2 V^{-3}$. Taking into account the $\omega^{-4}$ frequency dependence of the interband contribution (which plays a dominant role in the chosen frequency range) to $\sigma_3$ \cite{BaoLoh} we finally obtain the value
$\sigma_3 (\lambda= 565 \: nm) = -i 8.16 \cdot 10^{-23} A m^2 V^{-3}$ for the graphene nonlinear conductivity at $\lambda = 565 \: nm$.

\section{Graphene-assisted cavity response}
The field interacting with the cavity is transverse electric (TE) so that, with reference to Fig.1, it can be represented as ${\bf E} = e^{i k_0 (\sin \theta) x}A_y(z) \hat{\bf e}_y$ and ${\bf H} = e^{i k_0 (\sin \theta) x} \left[A_x(z) \hat{\bf e}_x +  \sin \theta A_y(z) \hat{\bf e}_z \right] / (c \mu_0)$, where $\omega$ is the radiation frequency, $k_0=\omega / c$ and $A_x$,$A_y$ are the independent field components. In order to exploit the transfer matrix approach, it is convenient to introduce the two component column vector $A = (A_x \: \: A_y)^T$ for representing the field. The incident (i) and reflected (r) waves for $z<-d-L$ are $A_i =  e^{ik_0 \cos \theta (z+d+L)} ( -\cos \theta \: \: 1)^T E_i $, $A_r = e^{-ik_0 \cos \theta (z+d+L)} (\cos \theta \: \: 1)^T E_r $ whereas the transmitted wave for $z>d+L$ is  $A_t = e^{ik_0 \cos \theta (z-d-L)} (-\cos \theta \: \: 1)^T E_t$, respectively. The connection between the fields incoming (in) and outcoming (out) from a homogeneous slab of thickness $\delta$ and dielectric permittivity $\epsilon$ is given by $A_{out} =  M_{(K,\delta)} A_{in}$ where
\begin{equation}
 M_{(K,\delta)} = \left( \begin{array}{cc} \cos(K \delta) & -i(K/k_0) \sin(K \delta) \\ -i(k_0/K) \sin(K \delta) & \cos(K \delta) \end{array} \right)
\end{equation}
is the standard slab transfer matrix and $K = k_0 \sqrt{\epsilon - \sin^2\theta}$. Therefore the fields at both sides of the graphene plane $z=0$ are given by
\begin{eqnarray}
\left( \begin{array}{c} A_x \\ A_y \end{array} \right)_{z=0^-} = M_{(K_G,L)} M_{(K_{Ag},d)} \left( \begin{array}{c} \cos\theta (-E_i + E_r) \\ (E_i + E_r)
\end{array} \right), \nonumber \\
\left( \begin{array}{c} A_x \\ A_y \end{array} \right)_{z=0^+} = M_{(K_G,-L)} M_{(K_{Ag},-d)} \left( \begin{array}{c} - \cos\theta E_t \\ E_t
\end{array} \right). \nonumber \\
\end{eqnarray}
By setting the matching condition at the graphene plane $A_y^+ - A_y^- = 0$ and $A_x^+ - A_x^- = c \mu_0 K_y$ (where $K_y$ is defined in Eq.(\ref{surface}) where $E_y=A_y^+=A_y^-$), one obtains two equations (which we do not report here) containing $E_i$, $E_r$ and $E_t$ and the complex factor
\begin{eqnarray}
\Omega &=& \cos \theta \left[\cos(K_{Ag} d) \cos(K_G L) - \frac{K_G}{K_{Ag}} \sin(K_{Ag} d) \sin(K_G L)\right] \nonumber \\
       &-& i \left[\frac{K_{Ag}}{k_0} \sin (K_{Ag} d) \cos(K_G L) + \frac{K_G}{k_0} \cos(K_{Ag} d) \sin(K_G L)\right], \nonumber \\
\Lambda &=& \left[\cos(K_{Ag} d) \cos(K_G L) - \frac{K_{Ag}}{K_G} \sin(K_{Ag} d) \sin(K_G L)\right] \nonumber \\
       &-& i \cos \theta \left[\frac{k_0}{K_{Ag}} \sin (K_{Ag} d) \cos(K_G L) + \frac{k_0}{K_G} \cos(K_{Ag} d) \sin(K_G L)\right], \nonumber \\
\end{eqnarray}
By eliminating the amplitude $E_r$, Eq.(\ref{BasicEq}) is readily obtained.

Once the amplitude $E_t$ is evaluated from Eq.(\ref{BasicEq}) at a given $E_i$ (and hence also $E_r$ is known), the above transfer matrix approach allows to evaluate the field in the cavity bulk. The optical intensity $I$ within the cavity is the magnitude of the Poynting vector
\begin{equation}
{\bf S} = \frac{1}{2} Re \left({\bf E} \times {\bf H}^*\right) = \frac{1}{2 c \mu_0} Re \left[ A_y \left(A_z^* \hat{\bf e}_x - A_x^* \hat{\bf e}_z \right)
\right].
\end{equation}
%


\end{document}